\documentstyle[12pt,epsf]{article}
\newcommand{\bea}{\begin{eqnarray}}
\newcommand{\eea}{\end{eqnarray}}
\newcommand{\beq}{\begin{equation}}
\newcommand{\eeq}{\end{equation}}

\newcommand\prl{Phys. Rev. Lett.}
\newcommand\prb{Phys. Rev. {\bf B}}

\begin{document}

\centerline{\bf \Large The Fractional Quantum Hall Effect}
\vskip 1 true cm

\centerline{Sumathi Rao \footnote{{\it e-mail
address}: srao@thwgs.cern.ch, sumathi@mri.ernet.in}}  
\centerline{\it Mehta Research Institute, Chhatnag Road, Jhunsi,}
\centerline{\it Allahabad 211 019, India.}

\vskip 0.5 true cm
\noindent {\bf Abstract}
We give a brief introduction to the phenomenon of the Fractional
Quantum Hall effect, whose discovery was awarded the Nobel prize in
1998. We also explain the composite fermion picture which describes
the fractional quantum Hall effect as the integer quantum Hall effect
of composite fermions.

\vskip 1.0 true cm

I would like to start my talk\footnote{Talk presented at the 
Prof. K. S. Krishnan Birth Centenary Conference on Condensed Matter
Physics, held at Allahabad
University, Dec 7, 1998} by mentioning that the the 1998 
Nobel Prize in Physics has been awarded  
for the discovery of Fractional Quantum Hall Effect 
to 

\begin{itemize}

\item{} Robert Laughlin - a theorist from Stanford University,

\item{} Horst Stormer - an experimentalist from Lucent technologies
(formerly Bell Labs), and

\item{} Daniel Tsui - an experimentalist from Princeton University

\end{itemize}

\noindent Their citation reads ~~- ~~
`` for their discovery of a new form of quantum fluid with
fractionally charged excitations''.

In this talk, I will try to describe  this new form of quantum
fluid and its fractionally charged excitations. However, since I am
speaking to a general audience and the phenomenon of the fractional
quantum Hall effect may not be familiar to all, I will start my talk
with a brief introduction\cite{FQHEBOOK} 
to the classical Hall effect, before I start
with the quantum Hall effect.

The Hall effect, discovered in 1879, is simply the phenomenon that 
 when a plate carrying an electric
current is placed in a transverse magnetic field, the Lorentz force
causes a potential drop perpendicular to the flow of current

\epsfxsize=5.5 in
\epsfysize=2.0 in
\begin{center}
\epsffile{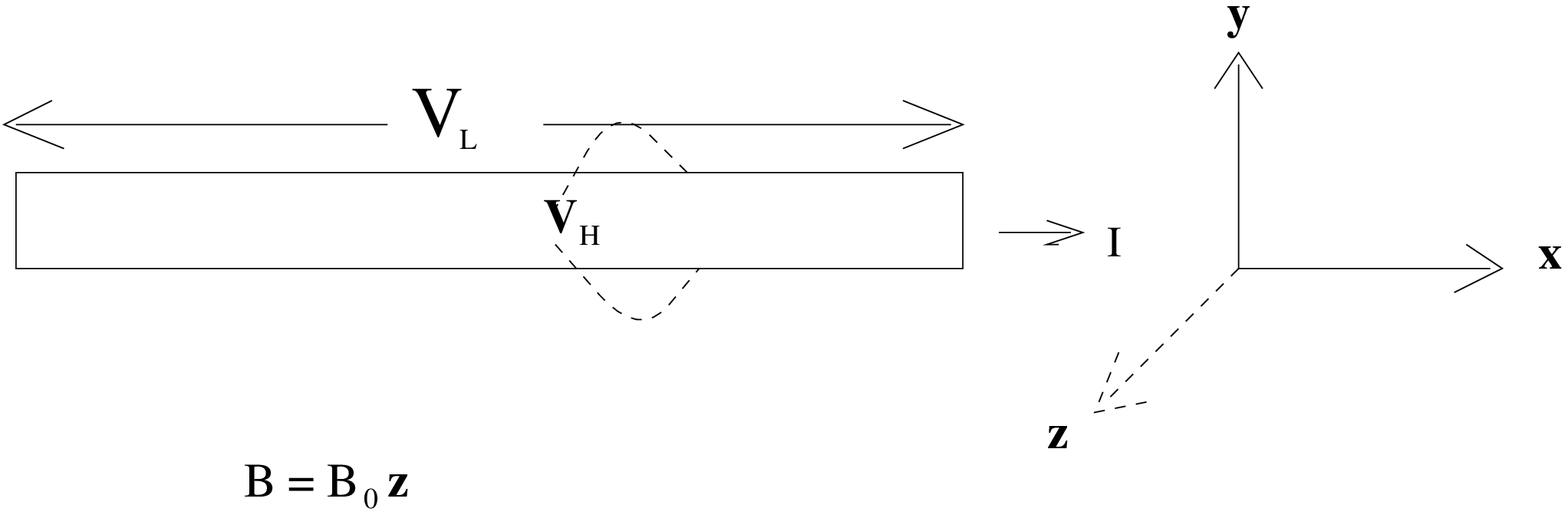}
\end{center}
\begin{itemize}
\item{\bf Fig. 1}
The Hall geometry.
\end{itemize}

This experiment is performed at room temperature and 
with moderate magnetic fields ( $\sim 1
~~Tesla$). If we measure the Hall resistance and plot it as a function
of the magnetic field, we get a straight line - $i.e.$, 
Hall resistance varies linearly with magnetic field.

\epsfxsize=3.5 in
\epsfysize=2.5 in
\begin{center}
\epsffile{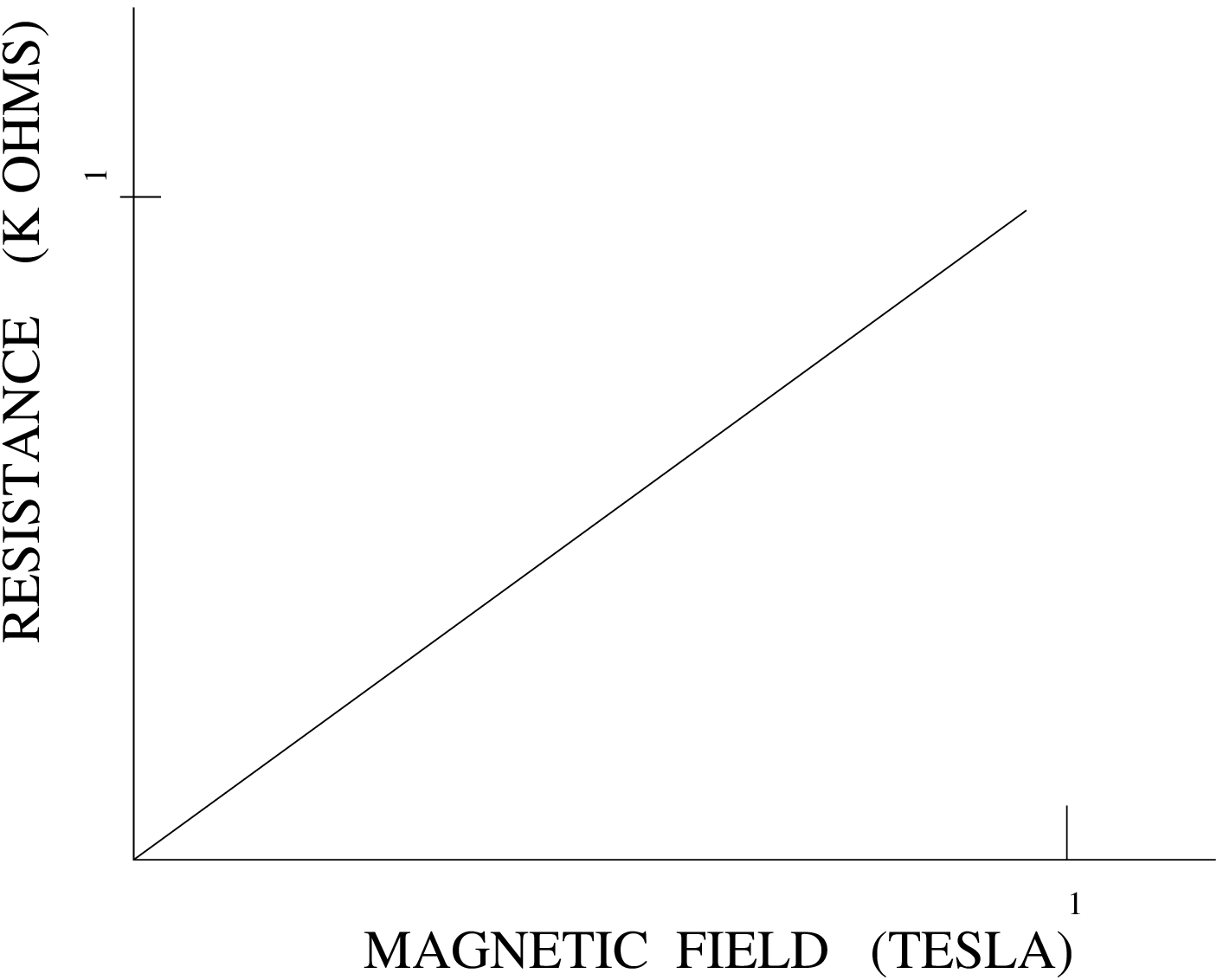}
\end{center}
\begin{itemize}
\item{\bf Fig. 2}
The linear Hall resistance  at moderate fields and room temperature.
\end{itemize}

Much later, in the early seventies, it was found that under certain
conditions, electrons could be made to  
effectively move only in two dimensions. This is achieved  by forming an 
inversion layer at the interface between a  semiconductor and an 
insulator 
($Si-SiO_2$) or between two semiconductors ($GaAs-Al_xGa_{1-x}As$).
In such a layer, at very low temperatures, (around $-272\deg C$), 
by applying an electric field perpendicular to the interface,
the electrons can be made to sit in a deep quantum well, which
quantises the motion of the electrons perpendicular to the interface.
Thus, the electrons  
are essentially constrained to move only in two dimensions.

In 1980, at very low temperatures ($1\deg Kelvin$) and at high
magnetic fields, (3-10 $Tesla$), Klaus von Klitzing 
discovered that the 
Hall resistance does not vary linearly with magnetic field, but varies in 
a 'stepwise' fashion, with the strength of the magnetic field. 
Even more surprisingly, the value of the resistance at these plateaux
was completely independent of the material, temperature, and other 
variables of the experiment and depended only on a combination of
physical constants divided by an integer -  $h/e^2\over n$!
This was the first example of quantisation of the resistivity.
(Note that for the Hall geometry, Hall resistance = Hall resistivity.)
In fact, the accuracy of this quantisation is so high, that it has led
to a 
new international standard  of resistance represented by the unit 
1 Klitzing = $h/ 4 e^2$ = 6.25 kilo-ohms defined as the Hall
resistance at the fourth step.
Note also  that where the Hall resistance was flat, the longitudinal 
resistance was found to vanish. In effect, the system was dissipationless  
and thus related to superconductivity and superfluidity.

For this discovery of the {\it integer quantum Hall effect 
(IQHE)}\cite{IQHE},
Klaus von Klitzing was awarded the Nobel prize in physics in 1982.

\epsfxsize=5.5 in
\epsfysize=3.5 in
\begin{center}
\epsffile{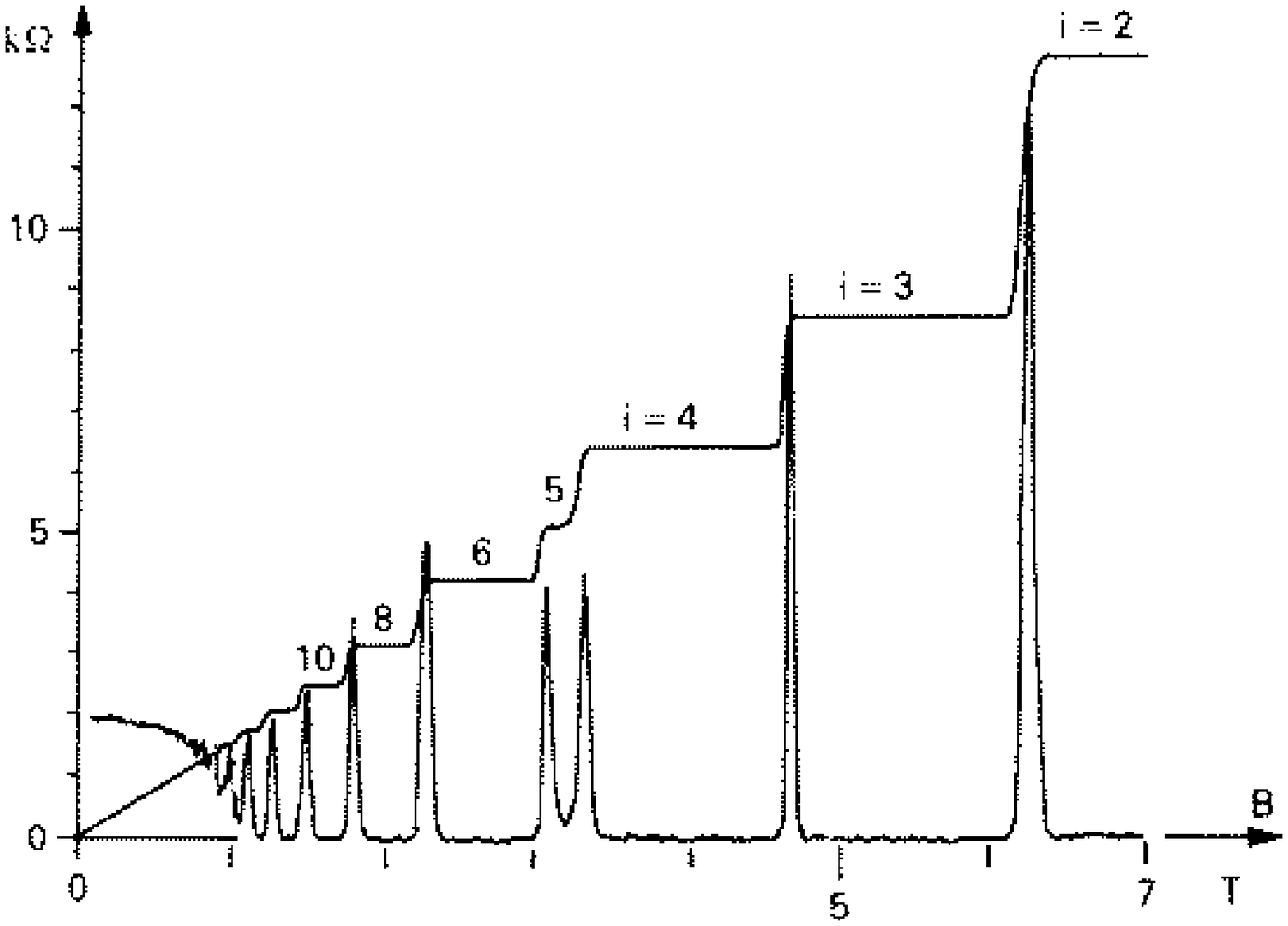}
\end{center}
\begin{itemize}
\item{\bf Fig. 3}
The Hall resistance varies stepwise with changes in magnetic field 
at high magnetic fields and low temperatures. The steps are quantised
at integer values of the filling fraction.(Kosmos 1986)
 
\end{itemize}

The IQHE can be easily explained using simple 
quantum mechanics of non-interacting
electrons in an external magnetic field. The Hamiltonian for the
system is given by 
\bea
H &=& \sum_i^N {({\bf p}_i - e {\bf A}({\bf x}_i))^2\over 2m}\nonumber \\
{\rm with}\quad {\bf A} ({\bf x}_i) &=& {B\over 2}(y_i, -x_i), \quad 
B = B{\hat z}.
\eea
Solving this Hamiltonian, we find the  energy eigenvalues 
$E_{n,k_y} = (n+1/2)\hbar\omega$, which are called 
Landau levels (LL) in terms of the cyclotron frequency
$\omega = eB/mc$. 
The Landau levels are degenerate, since they do not depend on the
$k_y$ quantum number.  The degeneracy of the Landau levels $\rho_B$  
(the number of states per unit area ) can 
be explicitly computed and is given by
$\rho_B = {eB\over hc}$. 

Let us define  a filling factor $\nu ={ \rho\over \rho_B}$ 
as the number  of electrons  per Landau
level. The filling factor can be thought of as a measure
of the magnetic field. Theoretical analyses are often
presented with the resistances as a function of the filling
fraction, rather than the magnetic fields.
In terms of the filling fractions, 
plateaux occur whenever $\nu = {\rm integer}$ or whenever
an integer number of Landau levels are fully occupied. 

Why does this happen? Let us see what happens as we increase
the density of electrons. As long as  states are available in the LL,
we can put more electrons into the level and the conductivity goes
on increasing (resistance decreases), but when a LL is full,
there exists an energy  gap to the
next available state in the next Landau level. But there exist 
localised states in the gap, due to impurities in the sample.
Hence, as the Fermi level passes through the gap, the localised
states gets occupied by the electrons and so do not contribute
to the conductivity. This is what causes the plateaux in the 
transverse conductivity, until the next Landau level is reached
and the same story is repeated.

To understand the extra-ordinary accuracy of the quantisation of the
resistance, one has to also realise the more subtle point that even
when some of the states in each LL get localised due to impurities, 
the conductance by the remaining states in that level is {\it as if
the entire Landau level was fully occupied}! In other words, the electrons
in the 
extended states move faster to compensate for the loss of the electrons
in the localised states. 

Another simpler hand-waving way to explain the  IQHE is to say that the
system is particularly stable when  an integer number of LL's filled. 
When we now add more particles, 
the system prefers to keep the average density fixed and
accomodate the extra particles as local fluctuations pinned by disorder.

Thus, IQHE is easily explained with just quantum mechanics of non-interacting
elctrons and the pinning of some of the states due to disorder.

In 1982, Horst Stormer and Dan Tsui\cite{TSUI} 
repeated the experiment with cleaner
samples, lower temperatures and higher magnetic fields ( upto $30
~~Tesla$). They found that the 
integers at which resistivity is quantized can now be replaced by fractions 
- 1/3, 1/5, 2/5, 3/7 $\cdots$.

\epsfxsize=5.5 in
\epsfysize=3.5 in
\begin{center}
\epsffile{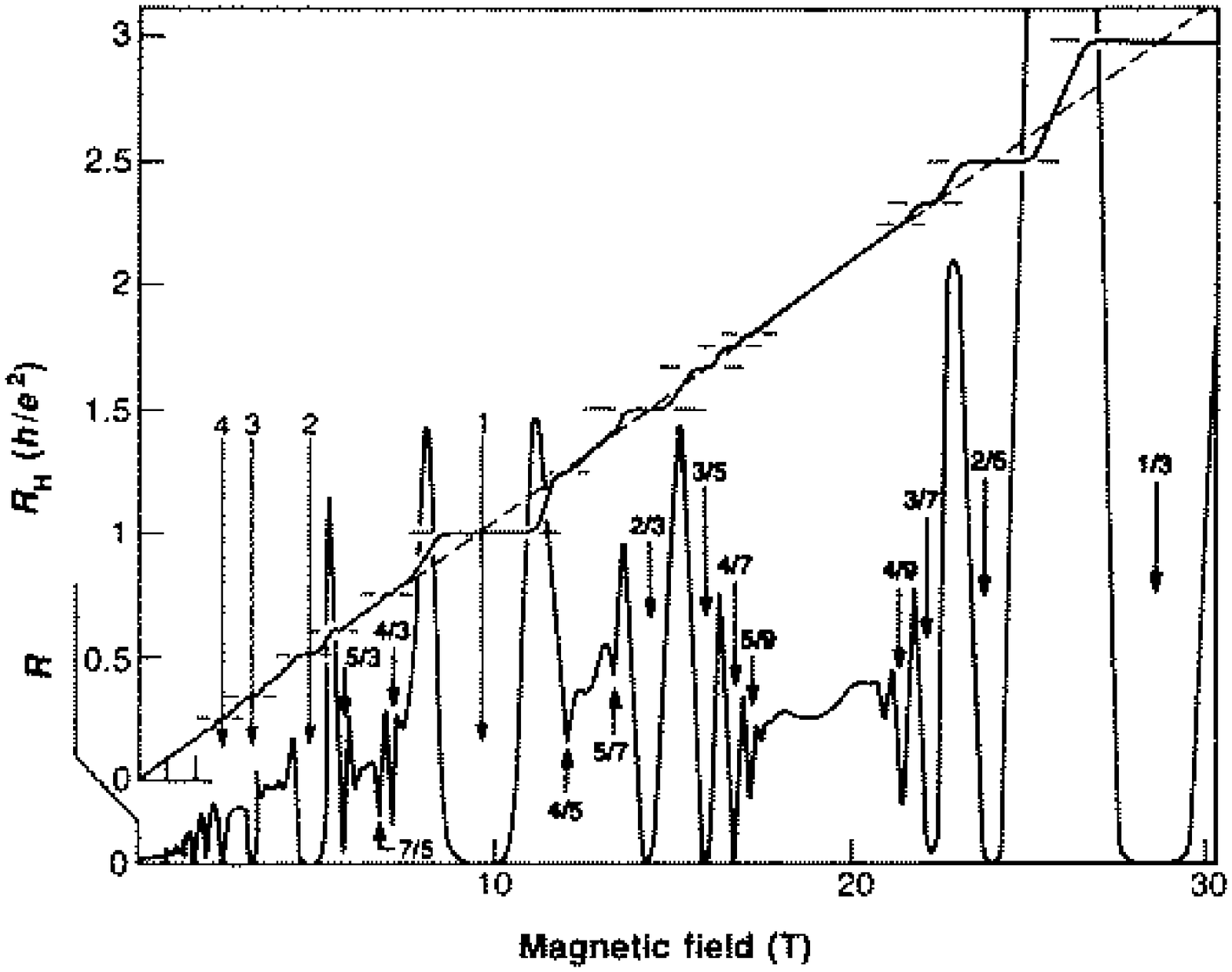}
\end{center}
\begin{itemize}
\item{\bf Fig. 4}
The Hall resistance varies stepwise with changes in magnetic field 
at even higher magnetic fields, lower temperatures and cleaner
samples. The steps are now quantised at fractional values of the
filling fraction. (Science 1990)
\end{itemize}

The QHE at these fractions could not  be explained by simple non-interacting
quantum mechanics, which says that 
at these fractions, the Fermi level is within the lowest LL
and so, the system is expected to be highly degenerate with no gap.
Without the gap, there is no stability and no possible 
explanation for the plateaux.

But this degeneracy is lifted because the electrons are interacting.
The Hamiltonian for interacting electrons is given by 
\beq
H = \sum_i^N {({\bf p}_i - e {\bf A}({\bf x}_i))^2\over 2m} +  
\sum_{i< j}^N {e^2\over |{\bf x}_i - {\bf x}_j|}. 
\eeq
Moreover, for fractions less than one, 
all the electrons are in the lowest LL. Hence, the kinetic energy is
completely quenched and the 
only relevant term in the Hamitonian is the inter-electron  Coulomb
repulsion. But the quenching of the kinetic term means that
$e^2$ is not small compared to anything. (For IQHE, on the other hand,
the potential energy $e^2/r_{av}$, where $r_{av}$ is the average
inter-electrons spacing, was small compared to the cyclotron energy
and could be neglected.) Hence, we  cannot use perturbation
theory and the 
problem is intrinsically one of strong correlations - a very hard problem.

Laughlin  in 1983  used a mixture of physical insight and numerical
checks to write down a wave-function - the by-now celebrated Laughlin
wave-function\cite{LAUGHLIN}   
\beq
\psi_L = \Pi_{i<j} (z_i-z_j)^{2p+1} e^{-\sum_i{z_i^2\over 4l^2}} \\
\eeq
- as a possible variational wave-function ( with
no variational parameters!) as an ansatz solution for the interacting
Hamiltonian.   
Here $z_i = x_i +i y_i$ is the complex position of the 
$i^{\rm th}$ particle and 
$l^2 = hc/eB$ is the  magnetic length. 

Using this wave-function, Laughlin could demonstrate  the following
properties -  

\begin{itemize}

\item{} The wave-function describes a uniform distribution
of electrons ( not random) $i.e.$, the number of particles within any
patch remains the same.  

\epsfxsize=5.5 in
\epsfysize=2.5 in
\begin{center}
\epsffile{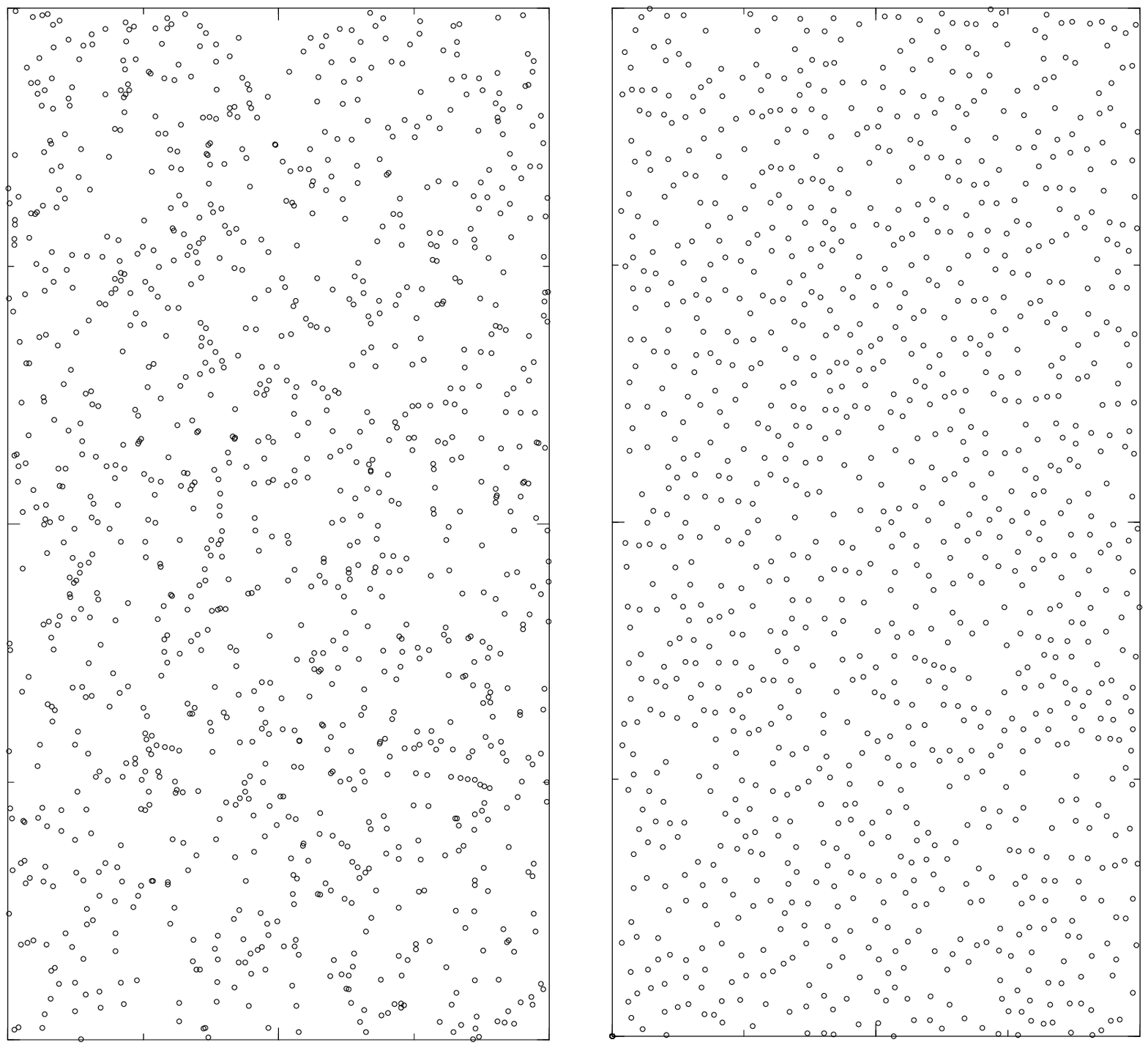}
\end{center}
\begin{itemize}
\item{\bf Fig. 5}
Comparison between a random distribution of particles (on the left) 
and a uniform
distribution of particles (on the right). 
\end{itemize}

Clearly, the uniform distribution (fluid-like) minimises Coulomb
energy  much better than the random distribution, which can have
patches with a large number of particles costing a large energy. In fact,  
$\psi_L$ was found to be very close to the exact ground state 
wave-function (calculated numerically) for small
systems.  

\item{} The state described by the wave-function is
incompressible. There exists a finite  energy gap
for  all excitations. This is a non-trivial point, since naively, one
expects a large degeneracy and instead, one  now finds that there
is a unique wave-function at these fractions with lowest energy and
all other possible wave-functions are less efficient in mimimising the
Coulomb energy and hence have higher energies. This  is related to the
fact that the Laughlin wave-function has multiple zeroes when two
particles approach each other, whereas Fermi statistics only needs a
single zero. These multiple zeroes are reponsible for uniformising the
distribution, which in turn, as seen in point 1) minimises the Coulomb
energy. 

\item{} Quasi-particle excitations over the ground state have 
fractional charges. 

\end{itemize}

This explains the citation which honours the scientists  
for their discovery of 
`` a new quantum fluid with fractionally
charged excitations ''

Why is the Laughlin wave-function ansatz so celebrated? Its fame 
lies in the fact that it is a correlated wave-function.
Normally, many-body wave-functions are Slater determinants of one-
particle wave-functions - 
$i.e.$, products of one-particle wave-functions appropriately
anti-symmetrised. For instance, the 
wave-function for one filled Landau level is given by 
\beq
\chi_1 = \left| \begin{array}{cccc}
                    1&1&\cdots&1 \\
                    z_1&z_2&\cdots&z_N \\
                    z_1^2&z_2^2&\cdots&z_N^2 \\
                    .&.&\cdots&.\\
                    .&.&\cdots&.\\
                    z_1^{N-1}&z_2^{N-2}&\cdots&z_N^{N-1}
               \end{array} \right| 
\nonumber
\eeq
which is a Slater determinant of single particle
wave-functions. Similarly, the two filled Landau level state involves
$z_i^*$'s as it involves the second Landau level, but it can still be
written as  a Slater determinant of one-particle levels in each of the
two Landau levels. 
But $\psi_L = \Pi_{i<j} (z_i-z_j)^{2p+1} e^{-\sum_i{z_i^2\over 4l^2}}$
cannot be written as sum of products of one-body wave-functions - it
intrinsically describes a correlated many particle state at a filling
fraction $\nu=1/(2p+1)$. 

Once, we have the result that at these fractions, the 
system is gapped, just like in the IQHE, it is easy to understand the
plateau formation by now having 
localised states in the intra-LL gap. Thus, using his wave-function,
Laughlin 
could explain the odd denominator rule, which simply comes from 
fermion statistics and FQHE at the 
fractions 1/3, 1/5, $\cdots$, 1/(2p+1). 
But more contrived scenarios (called the hierarchy picture) 
was needed to explain fractions like 2/5, 3/7,
$\cdots$.

In 1989, the next step in understanding the problem was taken 
by Jainendra Jain\cite{JAIN}. He  
identified the right quasi-particles of the system and called them
composite fermions.  
(There is no  guarantee that appropriate 
quasi-particles,  in terms of which any complicated strongly interacting
system appears weakly interacting, always exist, but the challenge is
to try and find them, if they do exist.
Phonons and magnons in lattices and spin models,
Landau quasiparticles in metals, 
Cooper pairs in  superconductors and  
Luttinger bosons(holons) in  one dimensional fermion models
are some examples.)
In terms of these quasi-particles, FQHE of strongly interacting
fermions is like IQHE of composite fermions.

The easiest way to understand his quasi-particles is pictorially. 
Let us 
measure magnetic field in terms of 
flux quanta per electron. 
IQHE at filling fraction $\nu = 1$ occurs when there is precisely one 
flux quanta per electron. FQHE, which occurs at higher magnetic fields
has more flux quanta per electron - $e.g.$, filling fraction $\nu=1/3$ 
corresponds to three flux quanta per electron.  Hence, we can depict 
IQHE as
\epsfxsize=5.0 in
\epsfysize=2.5 in
\begin{center}
\epsffile{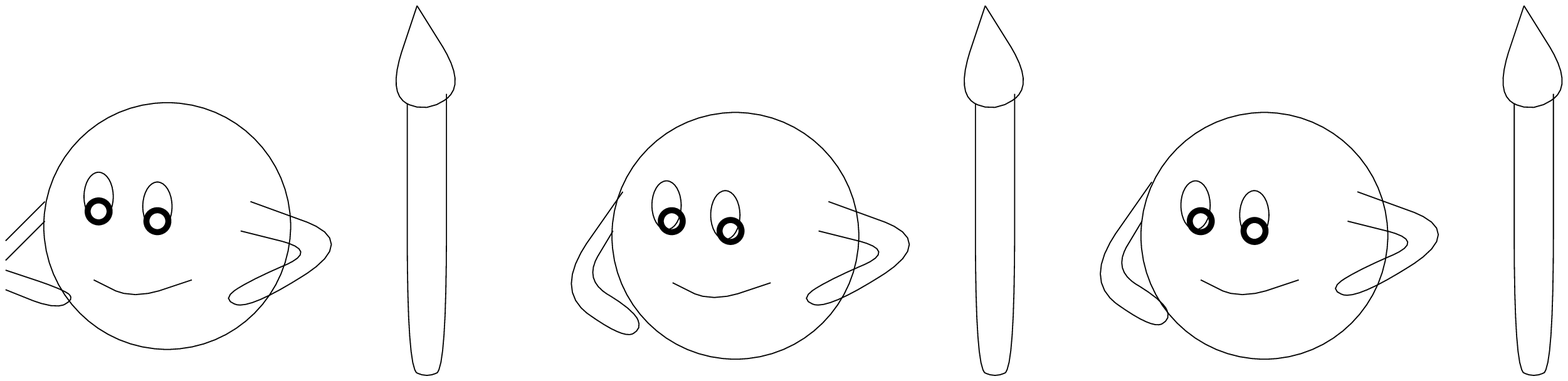}
\end{center}
\begin{itemize}
\item{\bf Fig. 6}
IQHE at $\nu=1$.
Electrons (depicted as balls) and flux quanta (depicted as tubes). On
the average, there is one flux quanta per electron.
\end{itemize}

\noindent and FQHE at $\nu = 1/3$ as 

\epsfxsize=5.5 in
\epsfysize=2.5 in
\begin{center}
\epsffile{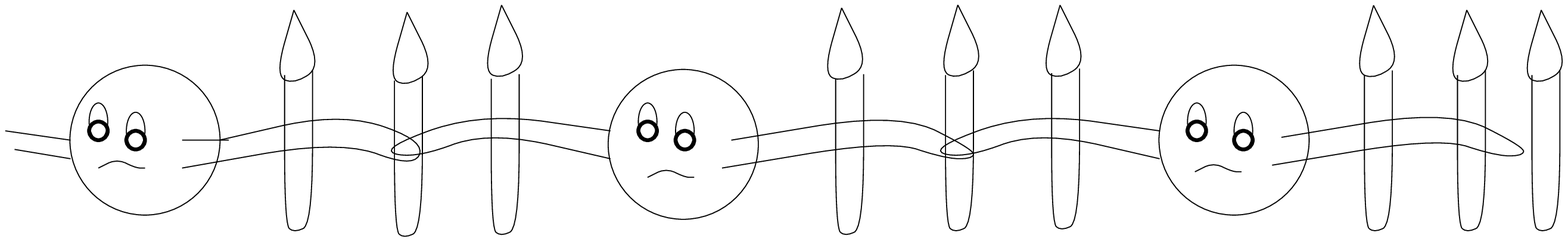}
\end{center}
\begin{itemize}
\item{\bf Fig. 7}
FQHE at $\nu=1/3$.
Electrons 'holding hands' implying strong interactions. On the
average, there are three flux quanta per electron.
\end{itemize}

\noindent Strongly interacting electrons are depicted as 'holding hands'!
Jain identified composite fermions as fermions with even number of flux
quanta attached - in this (simplest) case, two flux quanta are
attached. Hence, in Jain's picture,  
FQHE at $\nu =1/3$ is depicted as 

\epsfxsize=5.5 in
\epsfysize=2.5 in
\begin{center}
\epsffile{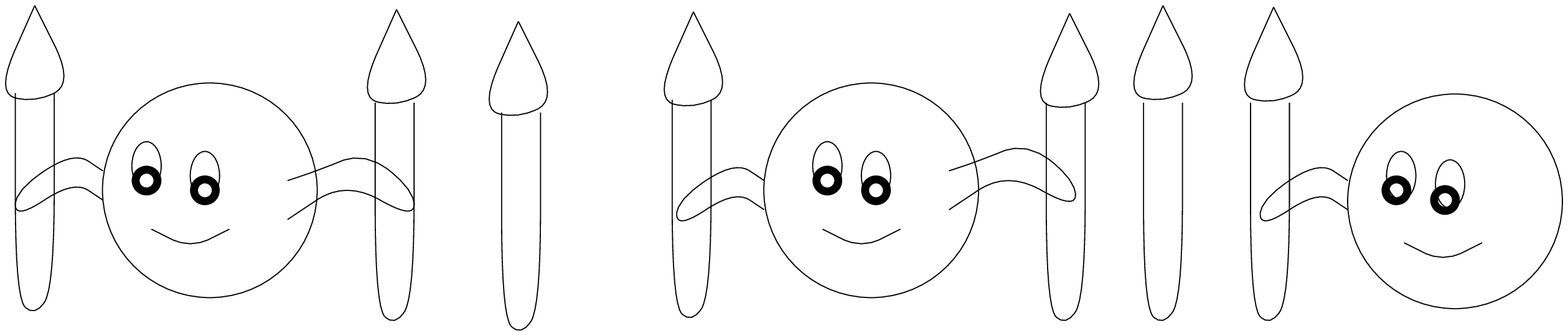}
\end{center}
\begin{itemize}
\item{\bf Fig. 8}
FQHE at $\nu=1/3$.
Composite electrons - electrons with two flux quanta attached - see on
the average, one flux quanta per composite electron.
\end{itemize}

\noindent Observe that composite electrons are no longer interacting!
They see one flux quanta per composite electron - similar to IQHE at $\nu
=1$ where electrons see one flux quanta per electron. Hence, FQHE is
analogous to IQHE of composite fermions. 

The explanation now for unique incompressible wave-functions at the
fractions is simply that they occur when  
composite fermion Landau levels are filled. Whenever an integer number
of composite fermion Landau levels are  completely filled, 
there exists a unique
ground state and a  gap to
the next level. This picture could not only accomodate all the
Laughlin fractions $\nu=1/(2p+1)$, but also all the hierarchy
fractions $\nu=2/5,3/7,\cdots$ at the same level.  
For instance, FQHE at $\nu = 2/5$ is just the IQHE of
composite fermions at $\nu =2$ and so on. 

Jain used this mean field picture to propose more general 
wave-functions than the Laughlin wave-function. He first 
rewrote the Laughlin wave-function as 
\bea
\psi_L &=& \Pi_{i<j} (z_i-z_j)^{2p+1} e^{-\sum_i{z_i^2\over 4l^2}}
\nonumber\\
&=& 
\Pi_{i<j} (z_i-z_j)^{2p}\chi_1 e^{-\sum_i{z_i^2\over 4l^2}} \nonumber
\eea
where $\chi_1$ is the wave-function of one filled Landau level. Then he
wrote the 
wave-functions for the other fractions of the form  $\nu = n/(2pn+1)$
(which are all the experimentally observed fractions) as  
\beq
\psi_{\rm Jain} = \Pi_{i<j} (z_i-z_j)^{2p}\chi_n e^{-\sum_i{z_i^2\over 4l^2}}
\nonumber
\eeq
where $\chi_n$ is the  wave-function of $n$- filled Landau levels. 
$n$ is the  filling fraction of  composite fermion Landau levels 
and the Jastrow factor 
$\pi_{i<j} (z_i-z_j)^{2p}$
turns composite fermions into fermions.

These wave-functions have been tested numerically and 
found to be very close to the exact
ground state. More interestingly, there now exists 
experimental evidence for composite fermions - the cyclotron orbit of the
charge carrier in FQHE has been shown to be determined by the effective
magnetic field seen by the composite fermion.

There is yet another way to understand incompressibility at the odd
denominator fractions. 
For the fraction $\nu =1/3$, consider fermions with three flux quanta
attached - a fermion with three 'hands' holding three flux quanta
depicted as 
\epsfxsize=5.5 in
\epsfysize=2.5 in
\begin{center}
\epsffile{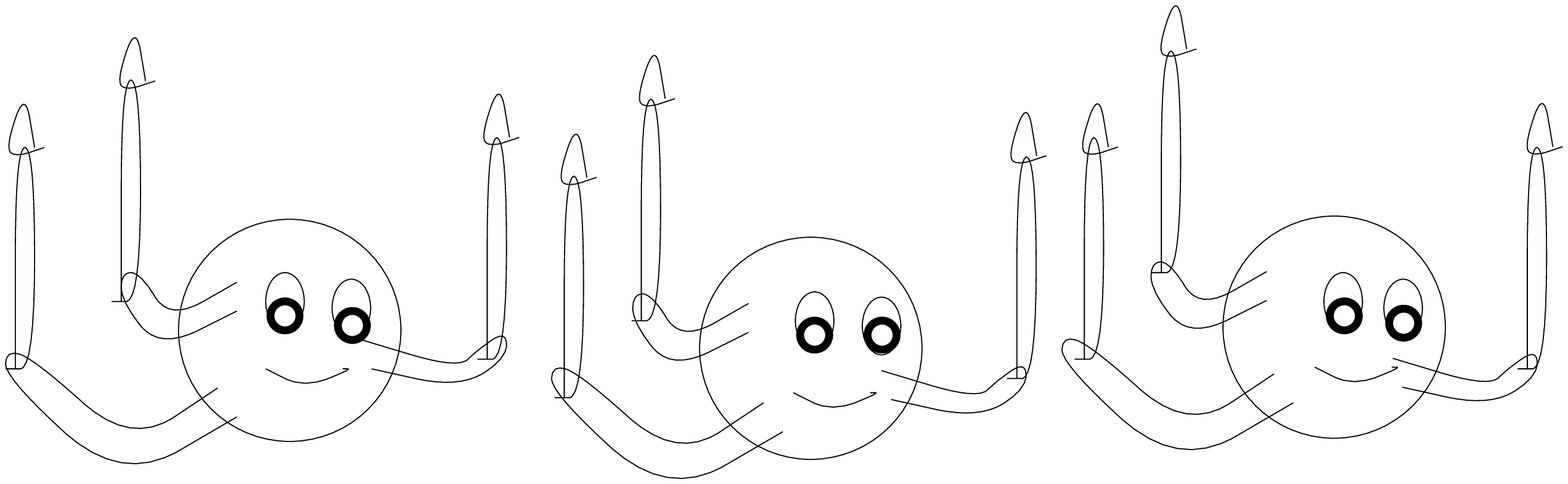}
\end{center}
\begin{itemize}
\item{\bf Fig. 9}
FQHE at $\nu=1/3$.
Composite bosons `holding' three flux quanta in zero field  Bose
condense.  
\end{itemize}
 
\noindent These  composite particles are now bosons  in zero field and Bose
condense. 
Hence, incompressibilty of the fermion system is equivalent to Bose
condensation of the composite bosons\cite{ZHANG}.   

Various explanations are possibe because in two dimensions, one can have
statistical transmutation and describe the same system in terms of
fermions, bosons or even anyons (particles with 'any' statistics).
However, composite fermions are really the appropriate quasi-particles
because they are the ones which are `weakly interacting'.

Finally, I will conclude by mentioning 
a few directions in which the subject is currently progressing
and give some examples of open problems. 

\begin{itemize}

\item{} Edge states at the edge of a sample of quantum Hall fluid.

Edge states form a chiral Luttinger liquid  and there have been 
several recent
experiments to probe edge physics. 

\item{} Double layer or multi-layer FQHE.

If the distance between layers is small, one can get 
new correlated electron states (with correlations between electrons
in different layers) as
ground states. 

\item{} FQHE with unpolarised and partially polarised spins

The usual FQHE assumes that the spin is completely polarised, so that
one is justified in working with spinless electrons, but there are
experimental situations where this is not true and one needs to
explicitly include the spin degree of freedom.

\item{} $\nu=1/2$ state.

The composite fermion picture yields 'free' fermions at
$\nu=1/2$. There has been a lot of interest both theoretical and
experimental in the study of this state which shows novel
non-Fermi liquid behaviour. 

\item{} Detailed calculations regarding the  widths of plateaux, transitions
between plateaux, effects of temperature, disorder, etc are yet to be
performed at a quantitative level.

\item{} At a more theoretical level, it is still an open problem to 
understand how microscopic Coulomb repulsions lead to the
formation of a composite fermion.

\end{itemize}

\end{document}